*Review*

# Environmental Remediation Applications of Carbon Nanotube and Graphene Oxide: Adsorption and Catalysis

**Yanqing Wang [1,2,3,*], Can Pan [4], Adavan Kiliyankil Vipin [2], Ling Sun [5,*] and Wei Chu [4]**

[1] Nihon Trim Co., Ltd., 2-2-22 Umeda, Kita-ku, Osaka 530-0001, Japan; y.wang@ipr-ctr.t.u-tokyo.ac.jp (Y.Q.W.)
[2] School of Engineering, The University of Tokyo, Bunkyo-ku, Tokyo, 113-0032, Japan; vipin@ipr-ctr.t.u-tokyo.ac.jp (A.K.V.)
[3] College of Polymer Science and Engineering, State Key Laboratory of Polymer Material and Engineering, Sichuan University, Chengdu 610065, China
[4] School of Chemical Engineering, Sichuan University, Chengdu 610065, China; 15053108110@163.com (C.P.); chuwei1965@foxmail.com (W.C.)
[5] Beijing Guyue New Materials Research Institute, Beijing University of Technology, 100 Pingleyuan, Chaoyang District, Beijing 100124, China; sunling@bjut.edu.cn
* Correspondence: y.wang@ipr-ctr.t.u-tokyo.ac.jp; sunling@bjut.edu.cn; Tel.: +81-3-5841-7461

**Abstract:** Environmental issues such as the wastewater have influenced each aspect of our lives. Coupling the existing remediation solutions with exploring new functional carbon nanomaterials (e.g. carbon nanotube, graphene oxide, graphene) by various perspectives shall open up a new venue to understand the environmental issues, phenomenon and find out the ways to get along with the nature. This review makes an attempt to provide an overview of potential environmental remediation solutions to the diverse challenges happening by using low-dimensional carbon nanomaterials and their composites as adsorbents, catalysts or catalysts support towards for the social sustainability.

---

## 1. Introduction

Along with the growing population, industrialization and urbanization, the lack of fresh and clean water is becoming a ubiquitous problem around the world [1-4]. Meanwhile, the shortage of water resources calls for efficient technologies for decontamination of wastewater, as well as the technologies for the seawater desalination. Hydrogen-dissolved water (electrochemically reduced water) studies performing high functionality are recently in hot topic [5,6]. Not to mention, over one half of the world population, mainly in Asia has to face a severe scarcity of safe water[7]. Environmental preservation has become a matter of major social concern. Although a lot of legislations have been imposed on effluent discharges, effective remediation processes are still highly desired to deal with non-readily biodegradable and toxic pollutants. It's well known dyes are usually used in textile manufacture, printing industry, and biochemical project, etc. Besides, that color removal with respect to organic effluents in discharged wastewater has been a compulsory measure and otherwise pertains to an illegal behavior[8]. The search for safe, effective and economic materials that can eliminate the current and future environmental contaminations from water is a scientific and technological issue of primary importance to the whole scientific community. Scientists around the world are devoting to find out effective materials either from natural or synthetic all for one purpose of environmental remediation.

Low-dimensional novel carbon nanomaterials e.g. carbon nanotubes (CNTs) and graphene oxide (GO) have been stimulating enormous interest in various scientific communities ever since their discovery. CNTs are a potential material used for a variety of applications because of its exceptional physical and chemical properties. The applications of CNTs are not limed to electrical, electronics, sensors, and thermal devices; moreover, they are emerging materials for environmental remediation [9-12]. CNTs are indeed a new class of material useful for environmental applications because of their cylindrical hollow structure, large surface area, high length to radius ratio, and hydrophobic wall and surface that can be easily modified [13].GO is one oxidized analog of graphene, recognized as the promising intermediate for obtaining the latter in large scale[14], since Brodie centuries ago first reported about the oxidation of graphite [15]. Three decades earlier, one atom-thin single layer of graphite was officially defined with the term graphene [16], structurally comprising $sp^2$ hybridized carbon atoms arranged in a honeycomb lattice, characterized by promising properties in terms of mechanical, electrical, and other[17-19]. Despite well-known properties, GO remains limited success in practical applications, mainly due to the difficulties in large-scale production of desired highly-organized structure[20]. While as one promising precursor, GO was much emphasized by academicians and by industry in the last decades [21,22], because it is readily exfoliated from bulk graphite oxide [23]. Such bottom-down strategy features of utmost flexibility and effectiveness arouse great interest in its practical applications.

Up to now, there are some reviews discussing about the recent advances in low-dimensional carbon nanomaterials such as carbon nanotubes, graphene oxide, and graphene derivatives in terms of wastewater treatment[4,11,12,24-33]. However, the effective design of multifunctional carbon nanomaterials via structural engineering, morphological control and component manipulation and get the utmost of them in the matrix during the application need to be paid more attention. Therefore, this review is intended to highlight the effect of functionalization and / or encapsulation of CNTs and GO in individual or complex states, facilitating the effectiveness of various environmental remediation techniques.

## 2. Carbon nanotubes (CNTs) based composite materials for water remediation

### 2.1 CNTs / functionalized CNTs as sorbents

CNTs were considered as a superior material for the remediation of a wide range of organic and inorganic contaminants comparing with conventional sorbents such as clay, zeolite, and activated carbon, because of their stronger chemical and physical interactions, rapid equilibrium, high sorbent capacity, and tailored surface chemistry[34,35]. Here we summarize the research achievements of CNTs in environmental remediation. We mainly addressed this section in two parts, concerning remediation of organic and inorganic pollutants.

Adsorption mechanisms of organic molecules on CNTs have been extensively studied. Multiple mechanisms are acting simultaneously such as hydrophobic interactions on the surface of CNTs, $\pi$-$\pi$ interactions, hydrogen bonds, and electrostatic interactions[36]. The adsorption of organic molecules on CNTs is mainly affected by the morphology and functional groups of CNTs and organic molecules. The adsorption also depends on the defects and active sites of CNTs[37]. Other environmental parameters that affected the adsorption are pH, ionic strength and dispersion state of CNTs[13]. The surfaces of raw CNTs are hydrophobic and show strong preference for adsorption of hydrocarbons (e.g., hexane, benzene, and cyclohexane), over alcohols (e.g., ethanol, 2-propanol). Surface functionalization reduces (usually) hydrophobicity, increases oxygen content, decreases specific surface area and, as a result, reduces adsorption of nonpolar hydrocarbons. Similarly adsorption of planar chemicals also decreases due to insufficient contact between CNTs and the chemicals. Adsorption properties as affected by CNT functional groups are shown in Figure 1.

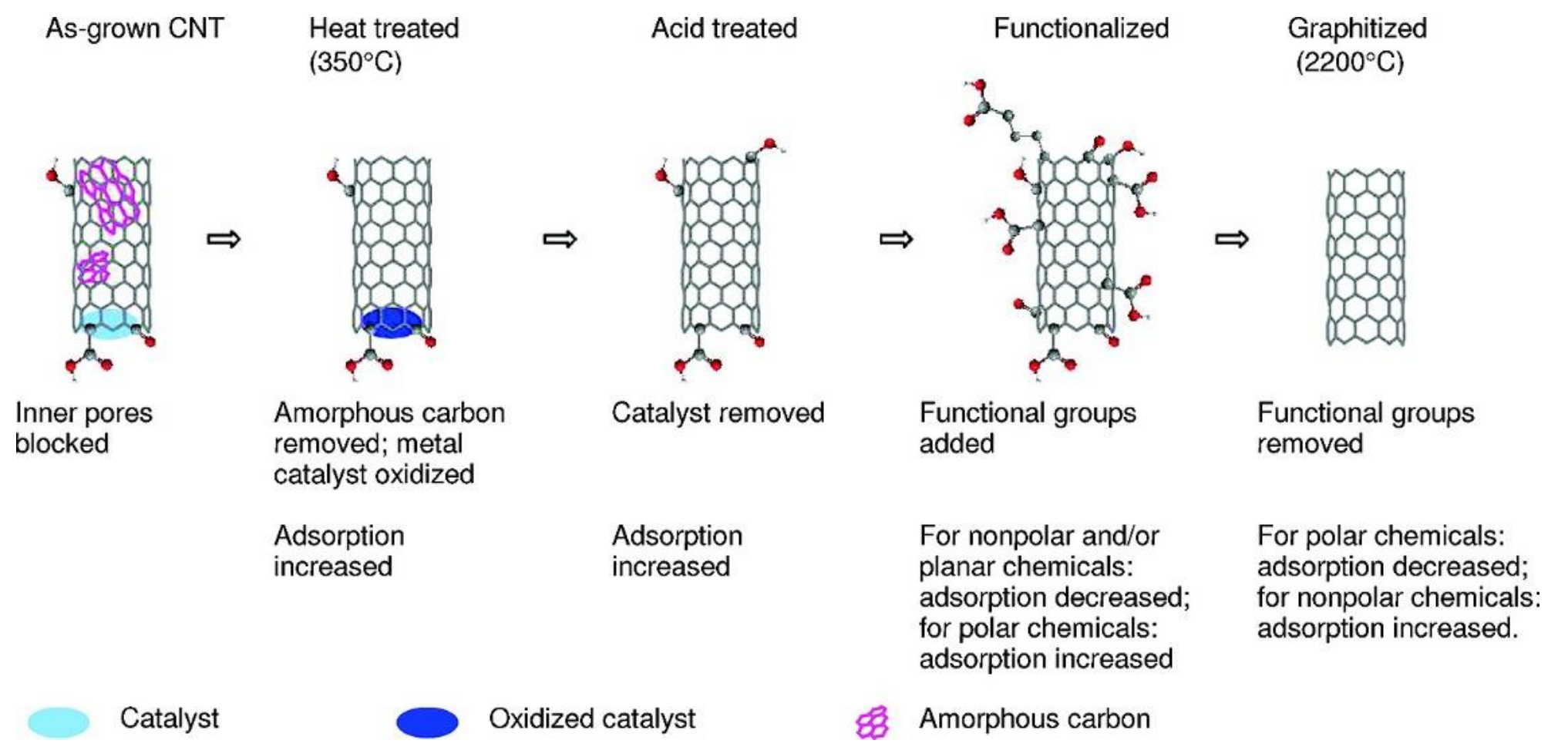

**Figure 1.** The general trend for the changes of CNT adsorption properties after different treatments. Reprinted with permission from [36]. Copyright (2008) American Chemical Society.

***Organic Dyes***

The first use of mono-dispersed CNTs as the active elements for the elimination of dyes was reported by Fugetsu et al.[38]. They developed a $Ba^{2+}$-alginate matrix constituting a cage, which holds the physically trapped multi-wall carbon nanotubes (MWCNTs) (Figure 2). The beads carry negative charges on their surface so that they were capable for the elimination of cationic dyes such as acridine orange (AO), ethidium bromide (EB), eosin bluish (EOB), and orange G (OG). The adsorption efficiency of MWCNTs encapsulated by barium alginate beads for AO, EB, EOB and OG were 0.44, 0.43, 0.33 and 0.31 μmol/mg, respectively. Chung-Hsin Wu examined the adsorption efficiency of MWCNTs for procion red MX-5B at various pH and temperatures. The saturation adsorption capacity was 44.64 mg/g without any modification of MWCNTs[39]. The adsorption of dye on CNTs decreased with an increase of pH or temperature.

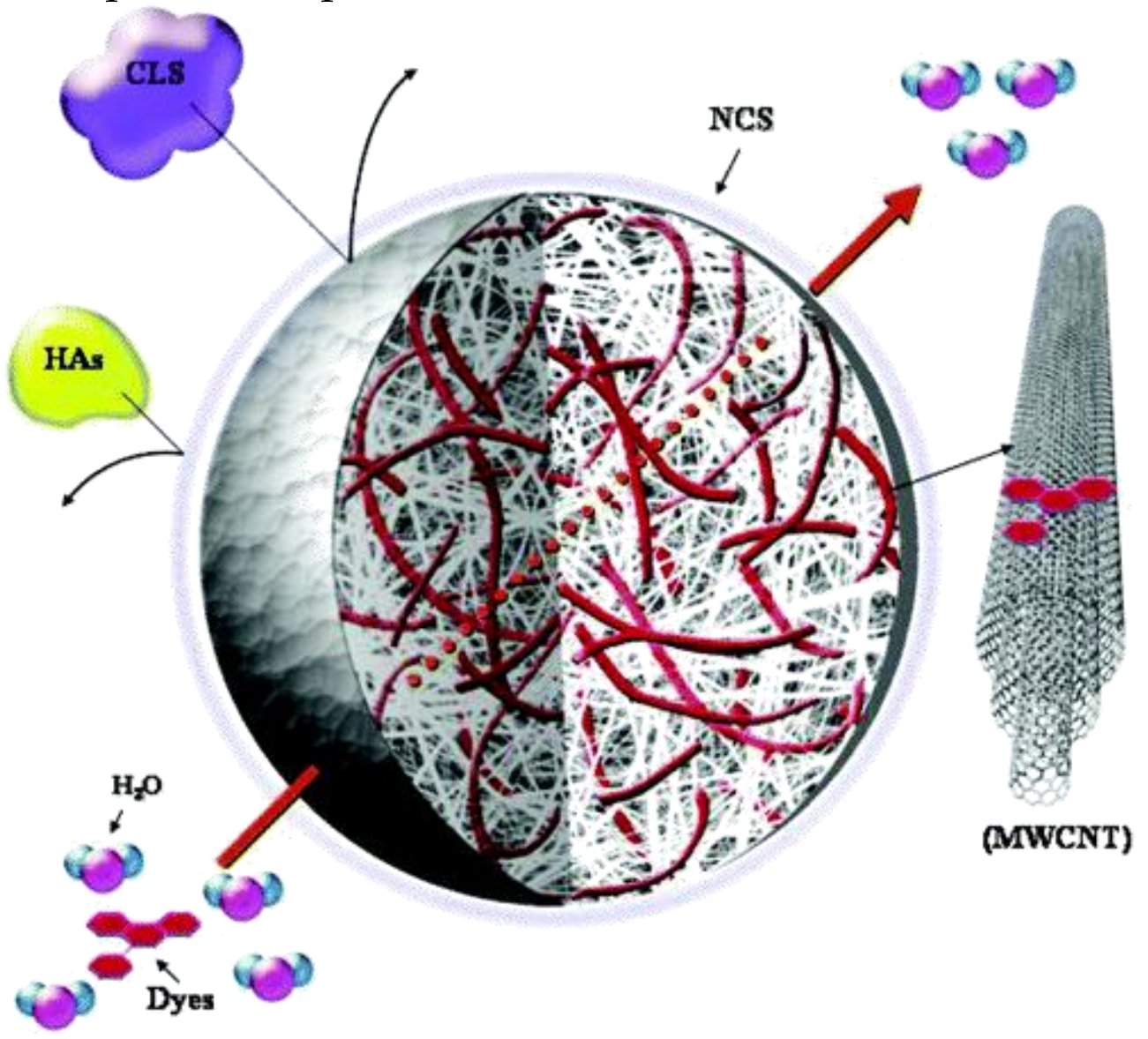

Figure 2. Schematic representation of the advantageous performances of the caged MWCNTs on dye adsorption. Reprinted with permission from [38]. Copyright (2008) American Chemical Society.

Ghaedi et al. did a comparative study of activated carbon and MWCNTs for efficient removal of Eriochrome Cyanine R (ECR). They studied the kinetics in detail, isotherms, and thermodynamics and observed that pH is important factor controlling adsorption. MWCNT in comparison with activated carbon (AC) has a higher adsorption capacity, and it reaches equilibrium in a shorter time. Physical forces, as well as ionic interaction, are responsible for the binding of ECR material[40]. Azo dyes are the most common synthetic dyes used in dye-works in the textile industry and the efficiency of MWCNTs as an adsorbent for removal of acid red 18 (AR18) from aqueous solution was determined. The research found that the Langmuir was the best fitted for experimental data with maximum adsorption capacity of 166.67 mg/g and the rate of adsorption was observed close to pseudo-second-order compared to other kinetic models[41].

Decontamination of cationic methylene blue from aqueous solution using carbon nanotubes was extensively investigated[42]. The adsorption capacity increased with temperature due to the endothermic nature of adsorption for CNTs and increasing mobility of dye molecules. They got a fantastic adsorption capacity of 132.6 mg $g^{-1}$ at 310 K[43]. The integration of MWCNTs with $Fe_2O_3$ nanoparticles has great potential application to remove neutral red (NR) along with methylene blue (MB). The adsorption capacities for MB and NR are 42.3 mg/g and 77.5 mg/g, respectively[44]. Iguez et al. researched the removal of MB together with Orange II (OII) anionic dye, from an aqueous solution by using MWCNTs. From the overall results, they concluded that MWCNTs could effectively remove both cationic and anionic dyes from aqueous solutions[45]. Gong et al. developed a magnetic multi-wall carbon nanotube nanocomposite for the removal of cationic dyes such as MB, NR, and brilliant cresyl blue (BCB) from aqueous solution [46]. Similarly, a magnetic polymer multi-wall carbon nanotube nanocomposite was developed by Gao et al. for anionic azo dyes removal[47]. Those composites were relatively easy to separate from the solution after adsorption using a magnet. CNTs based composites for adsorption of dyes were also reported, it is CNT–chitosan, CNT–activated carbon fiber (ACF), CNTs–$Fe_3O_4$, CNTs–dolomite, CNTs–cellulose, and CNTs–graphene [13,48,49].

***Other organic pollutants***

Chen et al. conducted systematic study on the adsorption of polar and nonpolar organic chemicals on CNTs. The adsorption increased in the order of nonpolar aliphatic < nonpolar aromatics < nitroaromatics; whereas, the adsorption affinity increased with the number of nitro functional groups[50]. The first systematic study on the adsorption of polycyclic aromatic hydrocarbons such as naphthalene, phenanthrene, and pyrene onto six carbon nanomaterials, including single-wall carbon nanotubes (SWCNTs) and MWCNTs, was investigated by Kun Yang group[51]. A linear relationship between adsorbed capacities and surface areas or micropore volumes of these carbon nanomaterials is obtained. Agnihotri et al. studied the adsorption of organic vapors on SWCNTs such as toluene, methyl ethyl ketone (MEK), hexane and cyclohexane. The relative adsorption capacities was in the order of toluene (maximum) > MEK > hexane > cyclohexane[52]. Similarly, SWCNT interactions with aromatic (benzene), aromatic and heterocyclic (thiophene), and nonaromatic (cyclohexane) molecules were investigated by Dennis Crespo and Ralph T. Yang. The strongest adsorption was observed for smaller SWCNTs, and the adsorption followed the order thiophene > benzene> cyclohexane[53].

Hazardous and carcinogenic trihalomethanes (THMs), the important (organic) contaminants has to be of primary interest.. However, CNTs were purified by acid solution and effectively used for the adsorption of THMs from the water[54]. The polyaniline (PANI) grafted onto MWCNTs by plasma-induced grafting technique were successfully developed by Shao et al. PANI/MWCNTs have very high adsorption capacities in the removal of aniline and phenol from a large volume of aqueous solutions, and PANI/MWCNTs can be separated and recovered quickly from solution by a magnet[55]. The authors also claimed that metal and metal oxide-based magnetic materials (such as $Fe_3O_4$) could be dissolved in acidic solution so that PANI/MWCNTs can be seen as promising material for the separation of organic pollutants from aqueous solutions. Peng et al. used pristine CNTs and graphitized CNTs as an adsorbent to remove 1,2-dichlorobenzene from water. The adsorption is fast, and it takes only 40 min to attain equilibrium. The adsorption capacity of pristine and graphitized CNTs is 30.8 and 28.7 mg/g, respectively. The pristine CNTs are better for adsorption of because they have a rough surface which made adsorption of organic molecules much more accessible[56].

Dioxins and related compounds such as polychlorinated dibenzofurans and biphenyls are highly toxic and stable pollutants. Richard Q. Long and Ralph T. Yang found that the interactions of dioxins with CNTs are much stronger than that with activated carbon[57]. Their experimental results show that CNTs are better sorbents for dioxin removal.

***Heavy metal ions***

Heavy metals include cadmium, chromium, zinc, and lead in water cause serious problem to the environment. CNTs have shown great potential as an attractive adsorbent for the removal of heavy metal ions from contaminated water[58]. The sorption capacities of metal ions by raw CNTs are very low but significantly increase when oxidized by $HNO_3$, NaOCl, and $KMnO_4$. The adsorption mechanisms are very complicated, and the contributions include physical adsorption, electrostatic attraction, precipitation and chemical interactions [58-60].

Chao-Yin Kuo and Han-Yu Lin have researched on the adsorption of aqueous cadmium (II) onto modified multi-walled carbon nanotubes. They found that microwave assistance involves oxidation using acids or oxidants. The research showed that The $Cd^{2+}$ adsorption capacity is increased by modifying the CNTs surfaces. The negatively charged surfaces of modified CNTs electrostatically favored the adsorption of $Cd^{2+}$ in MW/$H_2SO_4$/ $KMnO_4$-modified CNTs more than in MW/$H_2SO_4$-modified CNTs[61]. Ze-Chao Di et. al. developed ceria nanoparticles supported on aligned carbon nanotubes ($CeO_2$/ACNTs), a novel adsorbent for Cr(VI) from drinking water. The maximum adsorption capacity of $CeO_2$/ACNTs reaches 30.2 mg $g^{-1}$ at pH 7.0[62]. Chen et al. modified the surface of MWCNTs with polyacrylic acid and used for the europium adsorption. The incorporation of iron oxide magnetite enhances the separation and recovery after decontamination[63].

Several groups did a detailed study on the adsorption of lead ions using MWCNTs. They concluded that CNTs with a smaller diameter and higher oxygen content show greater lead adsorption ability [64,65]. SWCNTs and MWCNTs oxidized by NaClO for nickel removal from aqueous solution was reported by Chungsying Lu and Chunti Liu[66]. Similarly, $HNO_3$ oxidized CNTs was used by Changlun Chen and Xiangke Wang[67]. Another NaClO oxidized SWCNTs and MWCNTs used for the zinc adsorption was accomplished by Lu et al. Maximum zinc sorption capacity of SWCNTs and MWCNTs calculated by the Langmuir model were 46.94 mg/g and 34.36 mg/g, respectively[68,69]. In overall the heavy metal ion adsorption by CNTs follow roughly the order: $Pb^{2+} > Ni^{2+} > Zn^{2+} > Cu^{2+} > Cd^{2+}$. The application of CNTs was further extended to decontamination of water from radioactive cesium ions.

The sorption capacities by raw CNTs, especially the heavy metal decontamination, are low but significantly increase after functionalization, so most of the researches used oxidized CNT. In overall， CNT is an excellent material for the remediation of a wide range of organic and inorganic contaminants in contrast to many of the conventional sorbents. Further research on developing cost-effective CNT-related materials are recommended.

***Adsorption of cesium and strontium ions***

Removal of radioactive cesium and strontium from water has been an unsolved problem until today, being more urgent since the earthquake happened in Japan on March 11, 2011. Fugetsu et al. developed a quaternary spongiform adsorbent that contains Prussian blue, carbon nanotubes, diatomite, and polyurethane, which showed a high capacity for eliminating cesium both in laboratory studies and actual *in situ* Fukushima areas[70]. Low levels of radioactive cesium can be selectively adsorbed from sea water with a portable spongiform adsorbent that incorporates Prussian blue as the functional elements, and carbon nanotubes sealed diatomite as the working cavities (Figure 3). Thus-prepared complicated system was then permanently immobilized on the cell walls of polyurethane foam as the matrix. Such quaternary system shows that the elimination efficiency for cesium-137 was 99.93% in deionized water and 99.47% in seawater. Besides this study, Fugetsu group have been spending several years on developing caged approaches to overcome the difficulties associated with colloidal dispersion. We have tested organic polymers (like alginate) and porous inorganic particles (like diatomite) as potential caging materials to fabricate high performance adsorbents [71-77]. We intend to emphasize that highly-dispersed CNTs function not only as effective elements caged in surface negatively-charged gel micro-vesicles accomplished the affinity-based aqueous removal of typical ionic dyes, such as AO, ethidium bromide, eosin bluish and orange G beyond activated carbon and carbon nanofiber [78], but also as nano-thick network leak-proof coating to seal the Prussian blue-carrying-inside diatomite in the polyurethane composite, strengthened the resistance to radioactive irritation, and realized practical capture and transfer of radioactive heavy metal ions, such as cesium [70]. Previous researches, as mentioned above, having unveiled carbon nanotubes were promising for the purpose of decontamination towards various water-borne dyes and high-risk radioactive dissolved heavy metal ions[70,73,78-82].

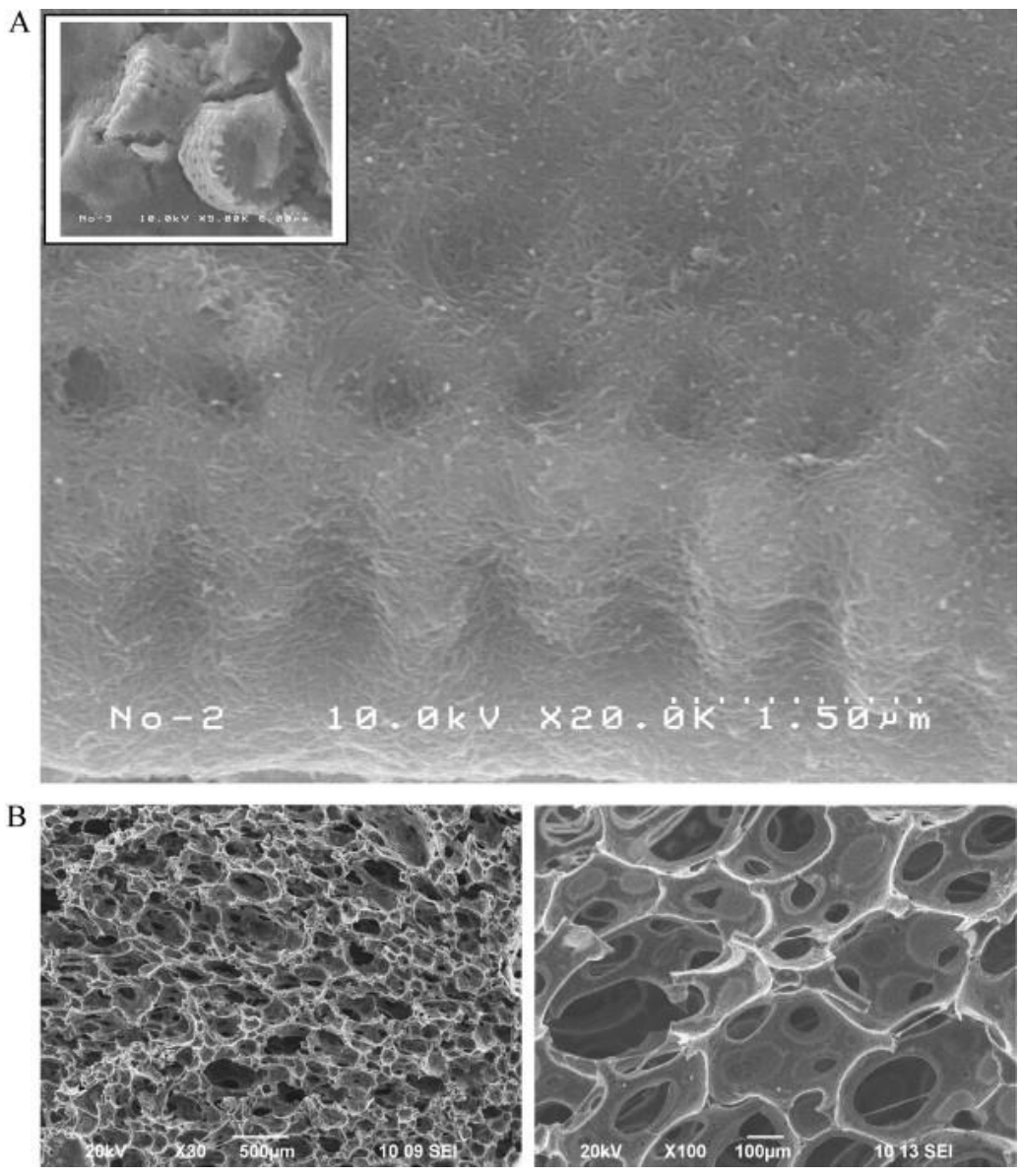

**Figure 3. (A)** Prussian blue was sealed into the cavities of the diatomite (upper, lower resolution). The diatomite surfaces were coated with highly dispersed multi-walled CNTs (upper, high resolution). The CNTs formed a continuous, interconnected network that prevented the diffusion of Prussian blue particles. (B) Representative SEM images of the quaternary (polyurethane polymer, CNTs, diatomite, and Prussian blue, PUP/CNT/DM/PB), spongiform, Prussian blue based adsorbent. Reproduced with permission from[70]. Copyright Elsevier, 2012.

Another kind of portable type of adsorbents (like beads) have been tried to adsorb the cesium ions and strontium ions from contaminated water by Fugetsu et al [80,82]. Prussian blue or its analogues encapsulating alginate beads reinforced with highly dispersed multiwalled carbon nanotubes were used for the studies on the removal mechanism and experimental adsorption of cesium and strontium ions, equipped with comparable kinetics models. Baiyang et al. and Vipin et al. used MWCNTs to enhance the immobilization of Prussian blue nanoparticle in polymeric cages [70,80,82]. They found that CNTs held PB more strongly inside the cage by forming a physical network over PB crystals which enhances the overall performance of adsorbent.

### 2.2 CNTs in catalysis reaction for water remediation

***Photocalytsis***

Even today, the wastewater purification is still a theme of large interest[11]. The so-called photocatalytic oxidation/reduction mediated by semiconductors has been discussed as a promising technology for the wastewater treatment in the scientific literature since 1976[83]. Various semiconductor materials such as $TiO_2$, $Fe_2O_3$, ZnO, etc. have been intensively studied since then [84-87]. However, their quantum efficiency is not high and the speed of ultraviolet photoresponse is not fast. Therefore, the development of modified $TiO_2$ with enhanced properties is needed to increase the photocatalytic activity for the organic pollutants[88]. Herein, carbon nanotubes are considered to be ideal catalyst carriers as to increase the quantum efficiency and extend the light adsorption region due to their huge specific surface area, remarkable chemical stability, unique electronic structure, nanoscale hollow tube property and good absorbability[27,89-93]. Herein, we summarized some typical studies of CNTs as catalyst support for water remediation in Table 1.

**Table 1. CNTs as catalyst support for water remediation**

| Materials | Mechanism | Pollutants | Major observations | Ref. |
|---|---|---|---|---|
| SWCNT/$TiO_2$ | Photocatalysis | Oil | Antifouling and self-cleaning | [94] |
| $TiO_2$/SWCNT aerogel | Photocatalysis | MB | High visible light photoactivity | [95] |
| NCNT/$TiO_2$ nanowires | Photocatalysis | MB | High degradation ability and wettability | [96] |
| MWCNT/$TiO_2$ | Photocatalysis | MB | High visible light absorbing | [97] |
| CNT sponge | Electrocatalysis | MB | Continuously high degradation efficiency | [98] |
| MWCNTs-OH-$PbO_2$ | Electrocatalysis | Pyridine | High degradation efficiency of pyridine | [99] |
| CNT thread | Electrocatalysis | Brine | Rapid deionization (2.78 mg g-1 min-1) | [100] |
| MWCNTs-silica aerogel | Adsorption | Oil | Superior oil adsorption capacity, 28.48 $cm^3$(oil)/g | [101] |
| NoCNTs | Catalysis | Phenol | Catalytic activity 57.4 times stronger | [102] |

Until now, a hybrid of $TiO_2$ and graphitic carbon (CNT, graphene, etc.) has gained tremendous research interest [103-106]. Ideal $TiO_2$/graphitic carbon hybrids reinforced with electroconductive and mechanically strong carbon nanotubes or a graphene backbone may possess an extremely large $TiO_2$/carbon interface that can facilitate electron-hole separation. Moreover, such interfacial hot spots may introduce a new carbon energy level in the $TiO_2$ band gap and thereby effectively lower the band gap energy[97,107]. Lee et al. have reported the N-doped CNT (NCNT)/$TiO_2$ core/shell nanowires for visible light-induced MB degradation by biomimetic mineralization of $TiO_2$ at the graphitic carbon surface[97]. They proposed that N-doped sites on the NCNT attracted Ti precursors and it promoted their condensation to form a highly uniform$TiO_2$ nanoshell with excellent coverage (Figure 4). Regarding to the NCNT/$TiO_2$ core/shell nanowires, the biomimetic mineralization yielded highly uniform coverage of a ~5 nm thick $TiO_2$ nanoshell over the entire length of the vertical CNTs because of the dense evenly distributed N-doping sites. The electroconductive and thermally stable NCNT backbone enabled high temperature treatment that optimized the corresponding crystal structure and properties of $TiO_2$ nanoshell. The direct contact of the NCNT surface and $TiO_2$ nanoshell without any adhesive interlayer introduced a new carbon energy level in the $TiO_2$ band gap, leading to the great visible light photocatalysis in MB degradation. Furthermore, the synergistic properties of core/shell nanowires also greatly enhanced the stimuli-responsive wettability. So this ideal multi-functional $TiO_2$/graphite-carbon hybrid nanostructure could facilitate a variety of artificial applications, including sensors, catalysts, energy storage and conversion.

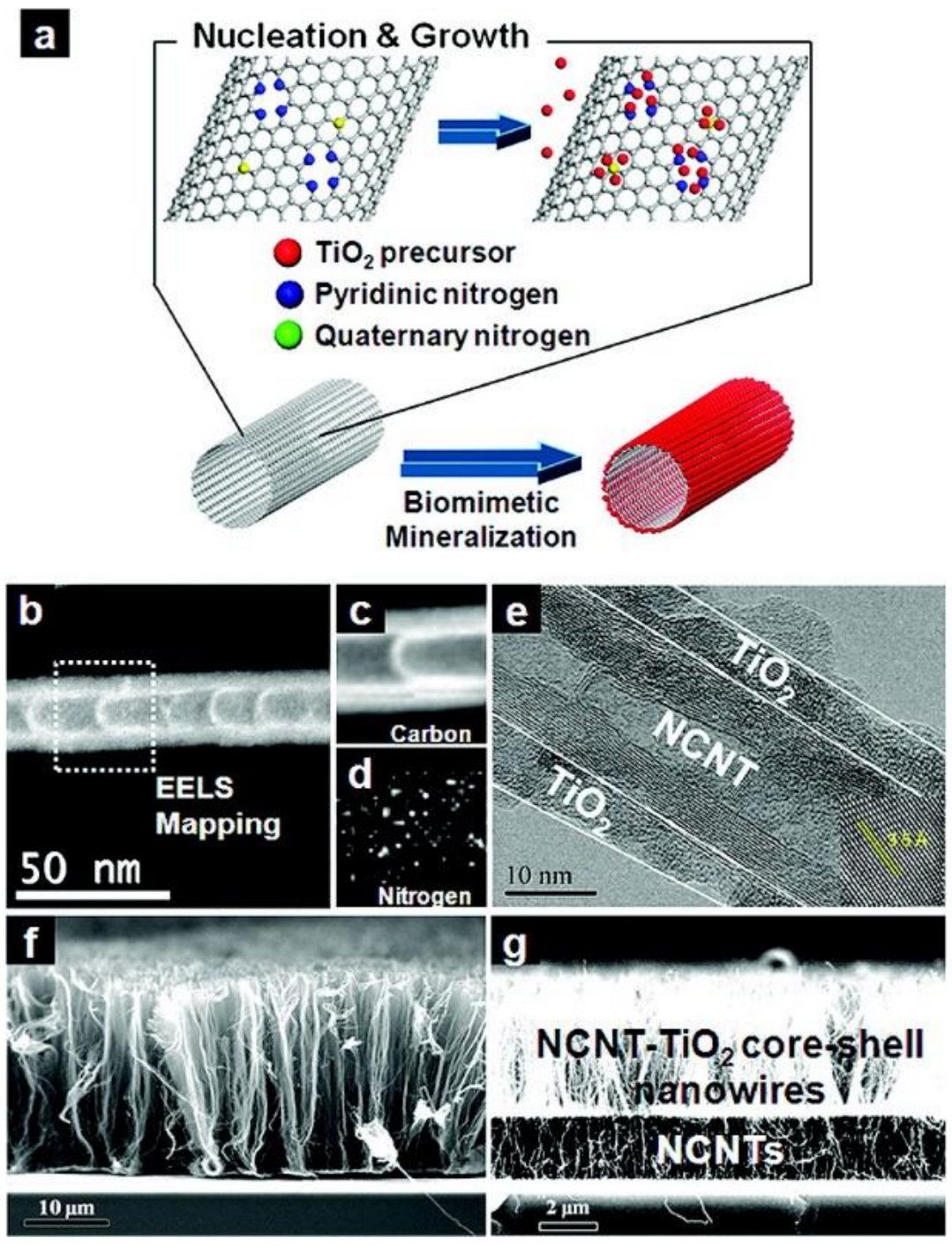

**Figure 4. (a)** Schematic illustration of biomimetic N-doped CNT (NCNT)/$TiO_2$ core/shell nanowire fabrication. Red, blue, and yellow colors indicate $TiO_2$ precursor, pyridinic N (NP), and quaternary N (NQ), respectively. N-doping sites act as nucleation sites. (b) ADF-STEM image of NCNT. EELS mapping shows (c) C and (d) N elements along NCNT. (e) TEM and (f) SEM images of NCNT/$TiO_2$ core/shell nanowires. The inset shows the lattice distance of the anatase phase. (g) NCNT/$TiO_2$ core/shell (top) and bare NCNT (bottom) heteronanowires. Reprinted with permission from [97]. Copyright (2012) American Chemical Society.

Furthermore, a lot of work are trying to make ultrathin and superwetting membranes to treat emulsified waste water produced in industry and daily life and also for purification of crude oil and fuel. Gao et al. reports on the preparation of an ultrathin and flexible film based on an SWCNT/$TiO_2$ nanocomposite network, with the aid of UV-light irradiation[94]. The ultrathin network film is superhydrophilic and underwater superolephobic, so it can effectively separate both surfactant-free and surfactant-stabilized oil-in-water emulsions with a wide range of droplet sizes (Figure 5). This robust and flexible SWCNT/$TiO_2$ nanocomposite films were prepared by coating $TiO_2$ via the sol-gel process onto an SWCNT ultrathin network film. The SWCNT/$TiO_2$ composite network film with adequate film thickness and nanoscale pore size shows high flux rate about 30000 $Lm^{-2}h^{-1}bar^{-1}$, and an ultrahigh separation efficiency achieved at 99.99%. Moreover, the characteristic of $TiO_2$ for photocatalytic degradation of organic compounds endows the SWCNT/$TiO_2$ film an excellent antifouling and self-cleaning properties even after being contaminated with oils. So the SWCNT/$TiO_2$ for removing oil from oil/water emulsions has potential at industrial scales. Besides, the freestanding $TiO_2$/SWCNT aerogel composites that have high visible light photoactivity were reported by uniformly decorating aerogels of individually dispersed SWCNTs with titania nanoparticles [95].

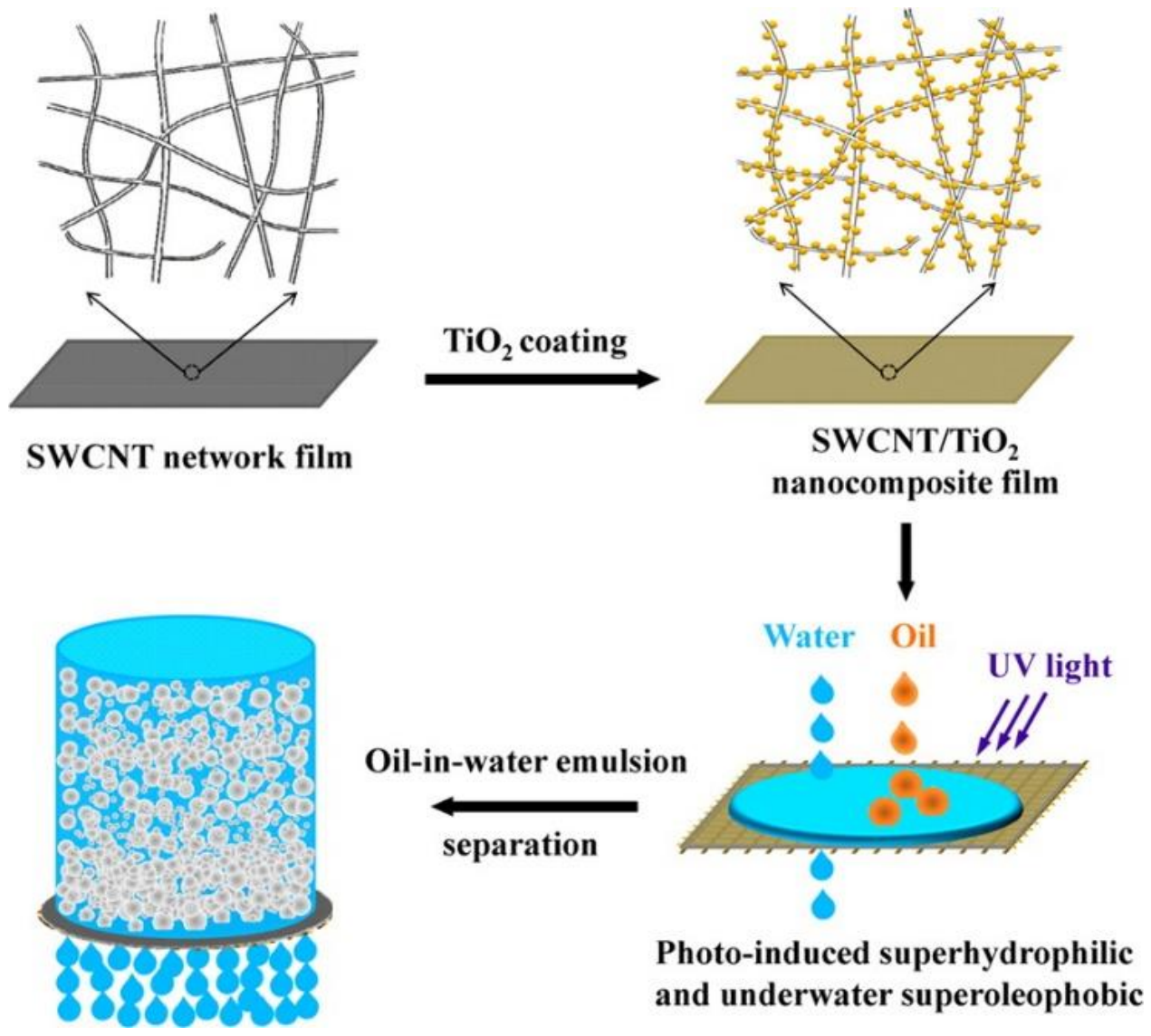

**Figure 5**. Schematic showing the preparation process of the SWCNT/$TiO_2$ nanocomposite film and for separation of an oil-in-water emulsion. Reprinted with permission from [94]. Copyright (2014) American Chemical Society.

***Electrocatalysis***

Electrochemical oxidation is another advanced oxidation process in the treatment of aqueous contaminations. It has attracted much attention because of its automation, fast reaction, environmental compatibility, and safety [108-111]. Some high surface area carbon-based materials such as carbon nanotubes have been extensively used to fabricate composites to adsorb pollutants, which are subsequently degraded by electrochemical oxidation [112,113].

Liu et al. have developed an affordable and effective electrochemical nano-sponge filtration device for water purification applications[98]. The assembly process was first conducted in an aqueous solution of CNTs facilitated by a surfactant, followed by a simple dying fabrication process for preparing high-performance conductive nano-sponge electrode. Both carbon nanotubes and polyurethane sponges play important roles in the design of the device. CNTs were incorporated to assist electro-oxidation, which can be used as a highly efficient pollutant degradation electrocatalyst and conductive additive to make polyurethane sponges highly accessible. Polyurethane sponges can be used as a low-cost, porous matrix to effectively carry these carbon nanotube conductors. The as-fabricated gravity fed device could effectively oxidize antibiotic tetracycline (>92%) and methyl orange (>94%) via a single pass through the conductive sponge under the optimized experimental conditions. The as-proposed water treatment technology might be an effective point-of-use device, as well as be scaled up and integrated into current water treatment systems to serve as a polishing step.

As a typical material, $PbO_2$ has been widely applied in electrochemical oxidation processes and exhibited high electrocatalytic activity[114,115]. In order to improve the oxidation activity of $PbO_2$ electrodes, Xu et al. prepared MWCNTs-OH-$PbO_2$ electrodes by adding hydroxyl modified MWCNTs in an electrodeposition solution for electrochemical degradation of pyridine[99]. MWCNTs-OH acted as electron barrier to hinder the successive growth of the big pyramidal $PbO_2$ particles, then the $PbO_2$ crystallites grew on the surface of pyramidal $PbO_2$ particles (Figure 6). This study revealed that the formation of small organic molecules by ring cleavage reaction and direct mineralization to $CO_2$ and $NO_3^-$ might be two potential pathways for electrochemical degradation of pyridine. Moronshing et al. recently reported a scalable approach to achieve efficient water purification by rapid capacitive deionization (CDI) with CNT-thread as electrodes[100]. Due to the synergism between electrical conductivity (~25 S $cm^{-1}$), high specific surface area (~900 $m^2g^{-1}$), porosity (0.7nm, 3nm) and hydrophilicity (contact angle~25°) in CNT-thread electrode, the extensive electrical double layer will come to rapid formation after contacting with water, and resulting in consequently efficient deionization.

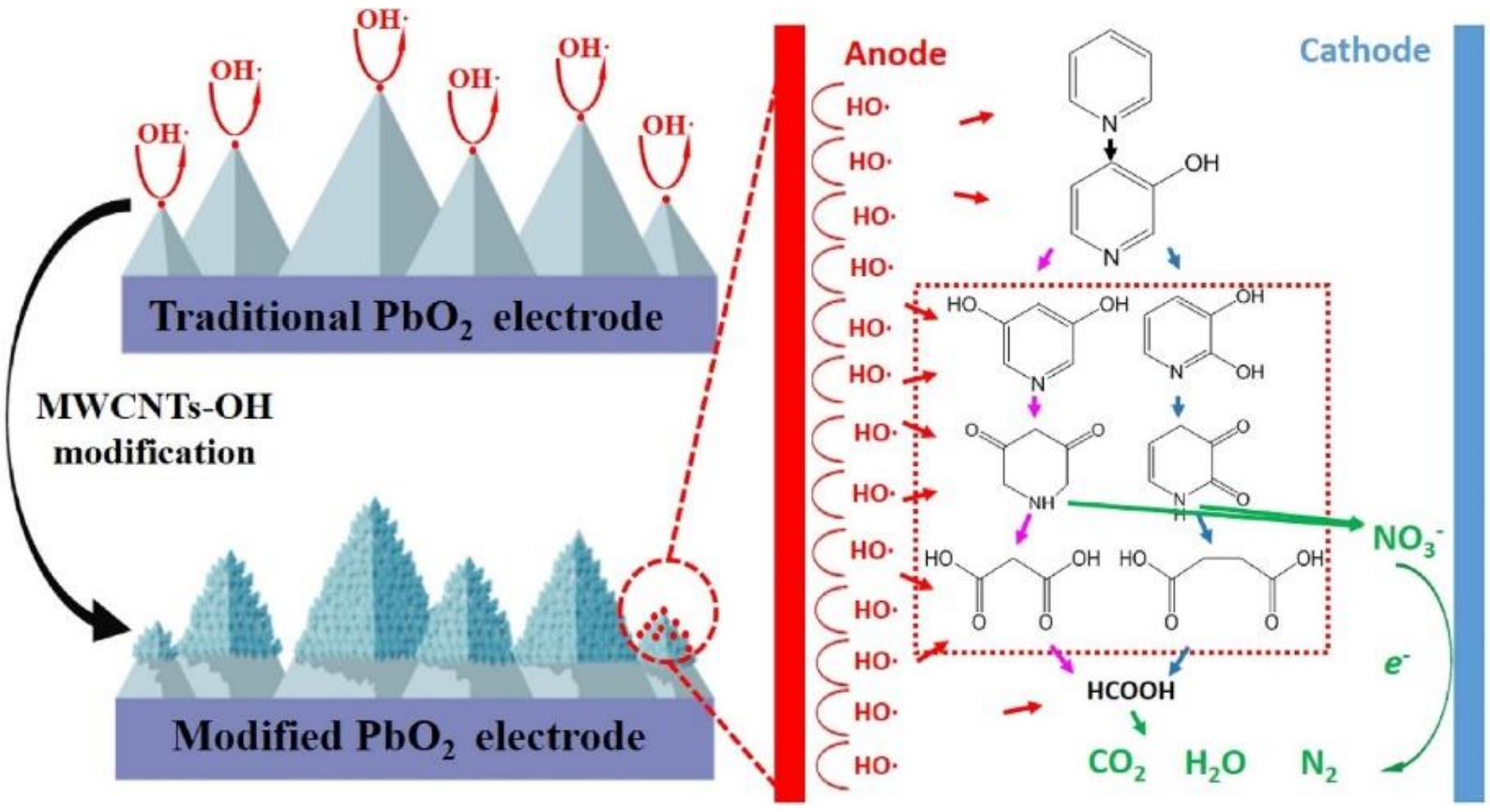

**Figure 6**. Hydroxyl multi-walled carbon nanotube-modified nanocrystalline $PbO_2$ anode for removal of pyridine from wastewater. Reproduced with permission from [99]. Copyright Elsevier, 2017.

***Other catalytic oxidation***

Recently, a novel oxidation technology based on nonradical activation mechanism of peroxydisulfate (PDS) by carbon nanotubes has received increasing attention, showing high reactivity towards phenol, bisphenol A, et al [116,117]. Unlike the case for other advanced oxidation processes in which free radicals are the keys for organic degradation, nonradical processes were also observed on the modified CNT samples upon nitrogen doping. Duan et al. prepared N doped carbon nanotubes (NoCNTs) as metal free catalyst by the chemical modification of SWCNTs with substitutional N incorporation into the rolled graphene sheets for phenol catalytic oxidation[102]. A comprehensive research for the mechanism of peroxymonosulfate (PMS) activation and the roles of nitrogen heteroatoms has been studied. Nitrogen heteroatoms in CNT played a significant role in phenol oxidation with PMS, which might greatly enhance both the radical and nonradical pathways that are beneficial to the phenol degradation. And the nonradical pathway accompanied by radical generation (•OH and $SO_4{\bullet}^-$) enable the NoCNT to have great stability (Figure 7). NoCNT presented an extraordinarily high catalytic activity for phenol removal by PMS activation with a 57.4-fold enhancement over the activity of NoCNT. Thus, this study provided insight to the role of N doped carbon nanotubes for enhanced catalysis and the designing of metal free catalyst with high catalysis performance and stability.

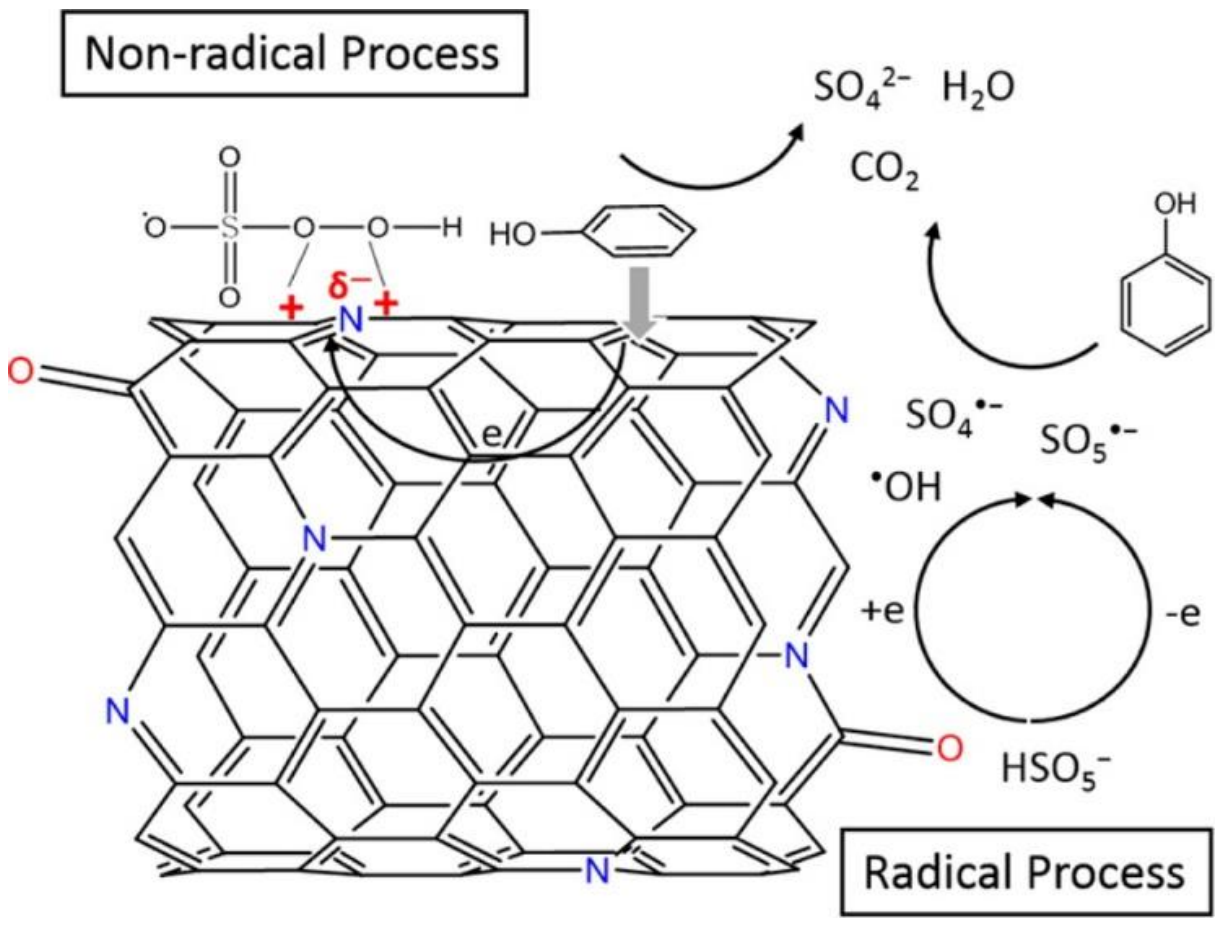

**Figure 7**. Mechanism of peroxymonosulfate (PMS) activation on N-Doped CNTs. Reprinted with permission from [102]. Copyright (2015) American Chemical Society.

## 3. Graphene oxide (GO) based composite materials for water remediation

Adsorption is one of the most important and basic technology as decontamination method. Its dramatic advantages over many other methods include ease of operation and comparatively low cost investment. Adsorption is the surface phenomenon whereby pollutants are adsorbed onto the surface of a material (adsorbent) via physical and/or chemical forces[118,119]. Along with the emergence and bloom of nanotechnologies, applications using carbon nanomaterials have exhibited great promise, especially the graphene derivatives [7,8,23,120-122]. As mentioned above, GO is two-dimension layered structure with large surface area and also rich in various oxygen-containing functionalities which include hydroxyl, carbonyl, carboxylic, and epoxide groups. Therefore, GO is rendered with high water-borne solubility in contrary to graphene, being easily mono-dispersed in the aqueous, unfolding its 2D structure and forming homogeneous colloidal suspension. Moreover, it can positively interact with various pollutants in molecular or ionic forms via the mechanisms, mostly known as the electrostatic interaction, pi-pi interaction, and hydrophobic interaction, etc.[120,121]. Therefore, GO has drawn much attention as a nano sized adsorbent [123].

Success of a typical adsorption relies on adsorbent providing active sites (imbalanced or residue force at their surfaces) and adsorbate capable of fast migration (Figure 8A). Thus, for activated carbon, the preparation has to go through an extra chemical /physical activation processing besides the carbonization of raw material to generate functioning sites for high performance adsorption. In contrast to bulky status of activated carbon, obtained graphene/GO as above mentioned is in nature of nano sizes, intrinsically with large surface area. Besides, the GO surface-grafted functionalities to large extent help unfold the sheets exposing to foreign contaminants as well as to resist GO's aggregation into flocculates in water, which may have great impact on the utilization performance.

As described, GO itself can straightforward act on contaminants, and adsorbs them at an extremely impressive level towards either ions or molecules. In our previous work[120], we selected AO as typical model contaminant. Firstly, it's a practical reagent in the printing and leather industries, and also a versatile fluorescent heterocyclic dye used in the biological research, for example, to observe RNA and DNA in living cells. The exposure to AO may genetically harm living beings as a negative consequence. Secondly, at that time, adsorptive research targeting such kind of dyes with graphene and GO were just a startup field, showing few reports on the influences of GO. And lastly, among those years, we had been paying much attention to carbon nanomaterials, especially the high-dispersion technology (Mono-dispersion Technology) dedicating to extensive applications [75,77], inclusive of environmental restoration, especially the water remediation. In last several years, the community of this field grew so fast that almost instantly graphene/GO had become the celebrities in carbon family. We are also involved in this journey, dedicating to shedding light on the great promise of chemical-exfoliation derived GO, by virtue of our solid nanocarbon research background.

### 3.1 GO as the sole sorbent

Table 2 summarizes the researches about the aqueous removal of dyes by GO-related adsorbents mainly in the recent decade. The section "as sole adsorbent" highlights GO itself functioning as the sole adsorbent. Notably, as to preparation of GO, most research referred to modified Hummers methods. As a consequence, there formed a variety of GO adsorbents yet with different terms, for example, GO, sGO,3D $GO_f$, $GO_p$. To some extent, this reflects a little dependence on the raw graphite type, or the expected structure of GO [120,121,124-150].

Similarly, highly-dispersed single-layered GO sheets or its derivatives were prepared by ourselves following the Hummers method [124,151,152]. Normally in a complete preparation, centrifugation as well as sonication is conventionally included in purpose of fully exfoliating quasi-stacked graphite oxide. And the sample was traditionally termed as GO. But for discrimination, modification is necessary on naming. Herein, the adsorbent from a modified Hummers' preparation with a 3-cycle sonication and yet without a centrifugation to screening the exfoliated graphite particles, was designated sGO [140]; the GO hydrogel adsorbent with a three dimensional (3D) structure and suffering reduction using sodium ascorbate was named as 3D RGO [131], which in fact was one less oxygenated derivative of GO; the adsorbent $GO_P$ designate to that from the specific graphite in powder form and the $GO_f$ did from the graphite flakes, while NGO defined the N-doped GO obtained by adding melamine of fixed amounts when the preparation of GO underwent [153]; the adsorbents GO-AG,FG,LG,VG were respectively used to describe GO from graphite with different graphitization degrees [142]. In general, GO can be facilely prepared via the Hummers method or other modified ways. However, through the same procedure, different groups including us can hardly assure the resultant GO sheets of each time being of absolutely identical characters due to the difference in terms of the types of graphite, some operational steps and/or the on-purpose treatment of structures. As to conducting the same research by different groups, the experiment is unavoidable to suffer from a big variation due to additional subjective factors, such as what graphite to choose, what modified Hummer method to use and so on. Thus, GO itself as adsorbent may already have non-negligible structural differences. This to a large extent explained the discrepancies of adsorptive performances in regard to the capacity (from thousands of microgram dye mass per gram GO mass to tens of), isotherms and kinetics, as shown in the Table 2.

Table 2. A summary of dye removal by GO materials

| | **Adsorbents** | **Dye** | **Isotherm type** | **Experiment (Calculation) mg g$^{-1}$** | **Kinetic type** | **Ref.** |
|---|---|---|---|---|---|---|
| As Sole adsorbent | $GO^{\#}$ | MB<br>MG | Langmuir | 220(351)<br>180(248) | - | [146] |
| | $HmGO^{\#}$ | MB | Freundlich | 387.9 | PSO | [144] |
| | $GO^{\#}$ | MB<br>RhB<br>CV | - | $199.2^{*}$<br>$154.8^{*}$<br>$195.4^{*}$ | PSO | [145] |
| | GO<br>SRGO | AO | Langmuir | 1382 (1428)<br>2158 (3333) | - | [120] |
| | GO sponge | MB<br>MV | - | 389<br>385.7 | PSO | [132] |
| | sGO | AO<br>MB<br>CV | - | 94.6<br>123.3<br>125.0 | PSO | [140] |
| | GO | AO | - | 229.8 | - | [154] |
| | GO | AO8<br>DR23 | Langmuir | 25.6(29)<br>14.0(15.3) | PSO<br>PSO | [125] |
| | GO | BR12<br>MO | Langmuir | (63.69)<br>(16.83) | -<br>- | [126] |
| | GO<br>AC<br>CNTs | MB | Langmuir | 240.65(243.9)<br>263.49(270.27)<br>176.02(188.68) | PSO<br>PSO<br>PSO | [130] |
| | 3D RGO | MB<br>RhB | Freundlich | 6.17*<br>9.18* | PSO | [131] |
| | GO | MB | Langmuir | (286.9) | PSO | [155] |
| | $GO_P$<br>$NGO_P$-1wt%<br>$NGO_P$-2wt%<br>$NGO_P$-3wt%<br>$GO_f$<br>$NGO_f$-1wt%<br>$NGO_f$-2wt%<br>$NGO_f$-3wt% | CR | Langmuir | (12.56)<br>(11.06)<br>(16.84)<br>(19.49)<br>(12.42)<br>(9.59)<br>(11.64)<br>(14.17) | PSO | [153] |
| | GO-VG<br>GO-LG | CB | Langmuir | 3071.47(3206.66)<br>3261.25(3414.92) | PSO | [142] |
| | GO-FG<br>GO-AG | | | 3404.80(3587.92)<br>3953.92(4248.79) | | |
| | GO | MB<br>BG | Langmuir | 927(476.19)<br>724(416.67) | PSO | [143] |
| | GO | MB | - | - | - | [149] |
| As Composite element | SA-GO-N<br>SA-GO-M | AO | Langmuir | 797(836)<br>1351(1420) | PFS<br>PFS | [121] |
| | RL-GO | MB | BET,Freundlich | (309) | PSO | [128] |
| | KGM | MO<br>MB | Freundlich | 51.6<br>92.3 | PSO | [139] |
| | 5wt%GOCB | MG | Langmuir | (17.862) | PSO | [150] |
| | MGSi | AO | Freundlich | - | PSO | [156] |
| | LI-MGO | AO<br>CV<br>OIV<br>GR | Langmuir | 62.08 (132.80)<br>37.88(69.44)<br>39.46(57.37)<br>147.8(588.24) | PSO | [135] |
| | GO-LCTS | MB | Langmuir | (402.6) | PSO | [155] |
| | GO/1-OA | MG<br>ER | Freundlich | 2687.56<br>1189.1 | PSO | [134] |
| | polyHIPEs/GO | MB<br>RhB | - | 1250.3*<br>1051.1* | PSO | [133] |

**Note:** - means the publication does not note; * the data is predicted by a kinetic model simulation; # the term GO used here indicates graphite oxide, but its preparation in fact followed the same Hummers method as those without this tag; RT means room temperature.

Most of the adsorptions of dyes onto sole GO agreed well with the Langmuir isotherm and followed the PSO kinetic model. In the early decade, only few reports discussed about nanocarbons in application of conventional dye removal since the high-cost disadvantage. As one early research, Bradder et al. [146] applied incomplete exfoliation form of GO (graphite oxide prepared by a modified Hummers method without post sonication) as an adsorbent for the removal of MB and MG from the aqueous. In contrast to graphite, they found the surface area of GO got little increased, while to the adsorption its surface oxygen functional groups played significantly important role. Intrinsic electrostatic attraction between GO and dyes resulted in a higher amount of the dyes adsorbed on the GO, and meanwhile a Langmuir-type monolayer isotherm.

To dig the superiority of merged carbonaceous nanomaterials, Li et al. (2013) [130] conducted the adsorptions of MB with nitric acid-treated carbonaceous materials, namely GO, activated carbon(AC) and CNTs. Due to the pi-pi electron donor-acceptor interaction and electrostatic attraction mechanism, all the adsorptions took on a Langmuir monolayer behavior with a pseudo second-order kinetic process. Differing in surface area accessibility, GO won CNTs and AC with the larger adsorption capacity normalized by the BET surface area.

Ghaedi et al. [143] researched the mixed adsorption of cationic dyes (MB and BG) by the GO prepared from a pre-oxidation-involved modified Hummers method. They found both the MB and BG removal were in a PSO kinetic process following a monolayer manner. They attributed the mechanism closely relative to the large surface area combining with the pi-pi electron donor acceptor interactions and electrostatic attraction between positively charged dye ions and negatively charged GO. Keeping searching more efficient adsorbents, Robati et al. [126] used commodity GO (atomic layer - at least 80%). It could remove the cationic BR12 and anionic MO within ~100 min from the aqueous at a pH ~3. It was found with the Langmuir monolayer adsorptive behavior through endothermic process. Interestingly, they observed the initial concentration of dyes affected oppositely on final adsorption capacity, namely an increase for the cationic BR12 while a decrease for the anionic MO if the initial dye amount increased.

Specifically to the anionic dyes, Konicki et al. [125] selected AO8 and DR23 as adsorbate for testing GO. Excluding hydrogen bonding, AO8 and DR23 are non-planar molecule hardly attaching to the skeleton of GO via formation of pi-pi stacking interactions due to spatial restriction, and thus, electrostatic attraction was probably the major contribution to the mechanism. As a result, the adsorptions presented more of a Langmuir model behavior than the R-P model and underwent in a PSO kinetic way. Meanwhile in another group, Jiao et al. [142] investigated the cationic blue with GO synthesized from four kinds of graphite with different graphitization degrees, designated as GO-AG (C/O ratio: 55.97), GO-FG (61.48), GO-LG (71.43), GO-VG (75.78). Likewise, all adsorptions came up with the same conclusion. Besides, highly disordered graphite (lower C/O) favored both the preparation and the adsorption capacity of GO. More and more recently, doping method emerged and endowed GO with more potential in water treatment. For instance, Yokwana et al. [85] prepared nitrogen-doped GO nano-sheets (NGO) using either graphite powder ($NGO_p$) or graphite flakes ($NGO_f$). The removal of the CR could achieve a maximum efficiency of 98%~99% for $NGO_p$ and 96%~98% for $NGO_f$ at the pH of 2. As concluded, there existed electrostatic interactions between the negatively charged (oxygen-containing) and positively charged (nitrogen-containing) groups on the NGO and the positively charged (amino and cationic azo-linkages) and negatively charged (sulfonic ($SO_3^-$) groups on the anionic CR dye molecules, possibly leading to the monolayer and PSO adsorptions.

From the perspective of GO, Sabzevari et al. [155] found GO prepared without sonication with the purpose of the removal of the cationic MB even suggested a relatively high monolayer uptake and took a PSO kinetic profile. More interestingly, He et al. [144] found their HmGO [157] prepared with no sonication too, led to the highest adsorption capacity for MB in comparison to other GO with sonication or from original Hummers method [158]. They found expansion of lamellar spacing, maintenance of lamellar structure and negatively charged oxygen-containing groups were favorable for the adsorption. The GO-MB chemical interactions led to the adsorption with a PSO kinetic process. A little earlier, Jin et al. [145] investigated the removal of cationic MB, CV and RhB dyes with the graphite oxide prepared by a modified Hummers method in which conventional mid- plus high-temperature reactions replaced a 12-hrs 70-°C sealed autoclave reaction. They speculated the dye molecules probably appeared in a head-to-tail form of aggregation when adsorbed in the layered structure, thus to some extent leading to different multilayer adsorption. With the chemical sorption nature, the adsorption still showed a PSO kinetic process.

It is worth of noticing that due to the homogeneity and large surface area, fully exfoliated, namely, single-layered GO could reach extremely high adsorption capacity. [120,134] However, the recovery of GO adsorbent becomes tedious and cost due to the operation inescapable of a prolonged ultrahigh centrifugation. To change the situation, a centrifugal vacuum evaporation method proposed by Liu et al. [159] was used to generate 3D GO sponges from suspensions and the sponges successfully removed over 98% of the MB and MV in two minutes, reflecting high capacity and fast rate. It was pointed out that the endothermic chemical adsorption mechanistically replied on the strong pi-pi stacking and anion-cation interaction with the activation energy of 50.3 and 70.9 kJ $mol^{-1}$. Tiwari et al. [131] synthesized uniformly mesoporous 3D RGO hydrogels by the reduction of a mixture of mono-, bi-, and tri-layer graphene oxides with large surface area using sodium ascorbate and simultaneous gelation. For the adsorption, they found such pi-pi stacking and anion–cation interactions dominated the mechanism, yet resulting in a Freundlich-type multi-layer adsorptive behavior with high removal capabilities for MB (100%) and RhB (97%) under initial concentrations of ~0.6 g $L^{-1}$.

In fact, GO was rather seldom studied to remove AO, unlike MB, MO, RhB etc. Only a few published contributions were found. Coello-Fiallos et al. [154] investigated the dye removal by the GO from natural graphite by modified Hummer's method. By analyzing the changes in the stretching vibrational bands of C-N and C-C, chemical interaction was corroborated between the AO and the functional groups of GO. The adsorptive equilibrium was found acquired within 1 hour with 40% removal efficiency. Afterwards, Coello-Fiallos et al. [140] continued the targeted study, including dye AO, with sonicated graphite oxide (sGO) prepared without centrifugation. Again, GO chemically adsorbed the cationic dyes AO more beyond CV and MB, and the processes followed the PSO model. Interestingly, Liu et al. [149] taking GO as a flocculant, observed that it could fast remove nearly all the MB from the aqueous and concluded that the MB cations were attracted by GO through the Van de Waals mechanism, other than frequently-reported electrostatic interaction as predicted by PIXEL energy contribution analysis in their research, and quickly congregated around GO in water. These conflicting mechanisms suggested further studies remain necessary, especially with multiple visual angles.

Prior to them, we ourselves targeted AO as a model contaminant. Through the simple adsorbent-adsorbate adsorption (Figure 8A, Figure 9A black line), a visible change of solution in color was obtained once the GO was added and demonstrated well elimination of AO from the aqueous phase. Likewise, modeling studies indicated all adsorptions behaved in a Langmuir-type monolayer manner. GO was found to have a maximal capacity of over 1400 mg $g^{-1}$, a record performance at that time. Even now, this is still remarkable as compared in the Table 2 [120]. At the same time, we noticed that all above researches were categorized as conventional one-step adsorption study, the final performance highly depended on the already-have functionalities or structures of the adsorbent, despite whether they are adsorption reactive or not. For full utilization of the 2D structure, we proposed a two-step adsorption (Figure 8B), of which GO was termed as SRGO (Figure 9A red line). The enhancement was realized via an additional in-situ slight chemical reduction to the GO that had reached the first equilibrium in the one-step adsorption. Amazingly, the transforming of its surface groups to more hydrogen-bonding active sites (pH<7) resulted in a remarkable increase of GO capability over 3000 mg $g^{-1}$ (Figure 9A red line), highlighting an extraordinary performance than GO itself. Meanwhile, such a two-step method did not alter the monolayer adsorptive manner (Figure 9B, Langmuir fitting). Importantly, this novel modification is of high repeatability, promising for emergency handling, which still pushing our team forward for more practical scenarios.

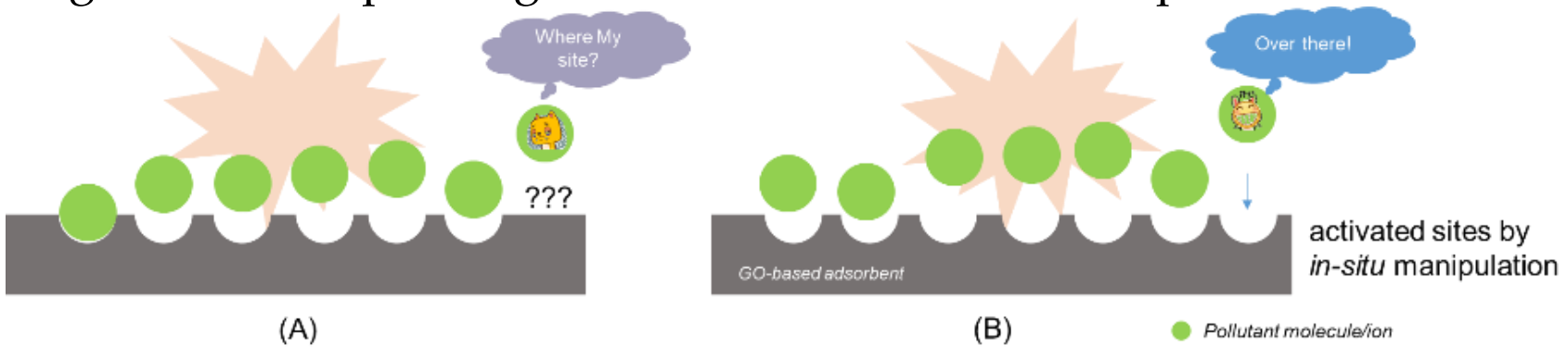

**Figure 8.** (A) The schematic diagram of conventional one–step adsorption; (B) Enhanced two-step adsorption: simultaneously generate new sites from inactive structures for enhanced capacity.

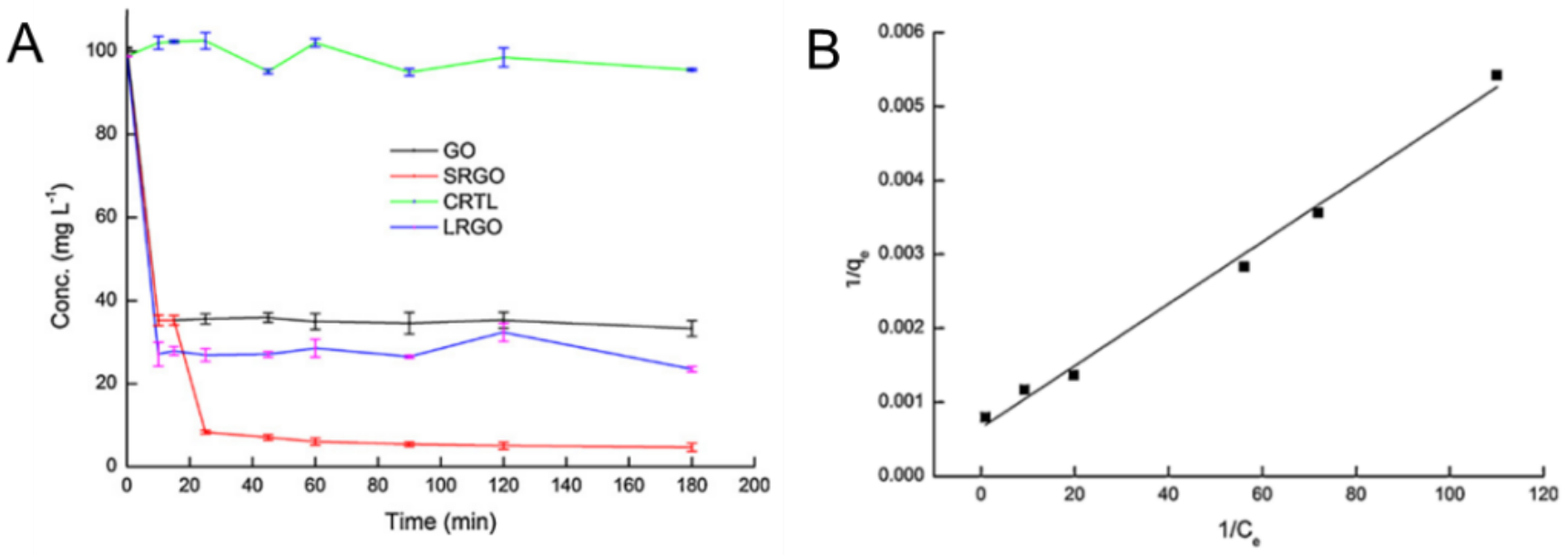

**Figure 9**. (A) Time-dependent adsorption of dye acridine orange by GO, in-situ reduced GO(SRGO), three-hour-long pre-reduced GO (LRGO) versus the blank group, AO ~100 mg/L, 50 ml; (B) Adsorption isotherms fitted by the Langmuir model: the reciprocal of equilibrium concentration versus that of equilibrium capacity. Adapted with permission from [120]. Copyright Elsevier, 2012.

### 3.2 GO as composited element

The section "as composite element" of the Table 2 summarizes some recent work using GO-composited adsorbents for typical dye removal. Notably, to the same purpose, GO-composited catalysts were also investigated. Frankly speaking, straightforward utilization of GO as the sole adsorbent still faces some problems, like inconvenient recovery and recycle, potential environment risk coming from its biotoxicity, and so on. To these ends, host materials were applied to composite GO inside. In the coming researches, main origins of novelty came from the merits inclusive of green preparation, high capacity, easy separation, recycle availability, etc. As a matter of fact, GO exhibited great potentials in an extensive scope for environmental applications and also such guest-host composites still suggested superiorities concerning the adsorptive capacity, batch kinetics and other indicators.

***Surface functionalization: GO as host material***

Conventional functionalization potentially risks from the fall-off of the loading materials from the host, influencing the working lifespan. To this end, Wang et al. [156] adopted a modified two-phase co-precipitation method and synthesized a magnetic calcium silicate graphene oxide composite adsorbent (MG) with improved joint stability. Consequently, such composite had selectivity of adsorption targeting AO other than CV and MB, ascribed to the higher electropositivity of AO in comparison to that of MB and less steric hindrance than the CV. The adsorption followed a pseudo-second-order kinetic model and yet in a Freundlich-type multi-layer manner. Wu et al. [128] examined a rhamnolipid-functionalized graphene oxide (RL-GO) composite prepared in one-step ultra-sonication as a cost-effective sorbent for artificial and real MB wastewater treatment. It contained abundant functional groups and a mesoporous structure and was insensitive to ionic strength variation when adsorption proceeded. It spontaneously and endothermically adsorbed the MB in a PSO and multilayer adsorptive manner (better fitting by both BET and Freundlich models), resulting from the synergetic mechanism of electrostatic attraction, pi-pi interaction and hydrogen bonding interactions.

Small organic molecules functionalized GO via non-covalent interactions are rarely explored for environmental remediation, due to the lack of stability in water of the resultant material though it is more versatile than covalent modification. Note that the 1-OA composite was formed from tetrazolyl compound 1 (1, 3-di(1H-tetrazol-5-yl) benzene) and octadecylamine (OA), having cooperative non-covalent forces to co-adsorb dyes and metal ions. Lv (2018) [134] brought this idea as reported into fabricating a two-component supramolecular 1-OA functionalized GO. As a consequence, in addition to easy solid-liquid separation and simple recycle, the modified GO/1-OA displayed a remarkable adsorption performance for pollutant dyes, BPA (endocrine-disruptor), Ciprofloxacin (pharmaceutical) and $Cu^{2+}$ in the single system or in their binary, ternary and quaternary pollutant mixtures. In the case of the cationic MG and anionic ER removal, the adsorptions suggested superior capabilities following the PSO kinetic mode and in a multilayer adsorptive behavior. Learned from these examples, we understood functionalization of GO would not weaken the whole removal efficiency for adsorbent but even got further enhancement. On the other hand, structure fragility remains as a fact what we worried about.

***Crosslinking: GO in host materials***

Several advantages of this strategy make it especially attractive. First, the fabrication goes always with a relatively simple and non-toxic route. Second, the adsorbent shows stable structure configuration and good adsorptive performances. Third, the adsorbent can be easily isolated by hand from the targeted solution and regenerated by some simple methods, for example, re-immersing in reactive solution [160], such as $Ca^{2+}$, $H^{+}$ and so on.

From this point of view, polysaccharides are commonly seen as green and economic biomaterials around us, including alginate, cellulose, chitosan and so on. With targeting the removal of AO and considering the porosity enrichment by inserting GO into many other matrices, we prepared GO homogeneously-loaded millimeter-sized beads (Figure 10A-D) [121] with biocompatible sodium alginate cross-linked by either $Ca^{2+}$ or $H^{+}$. These beads were characterized with a remarkable increase in pore structure and therefore exposed more functioning sites; in other word, they possessed better performances by contrast to that of those references. The ion exchange (Figure 10E) based electrostatic interactions as the main mechanism was found to play a vital role and led to the Langmuir-type adsorption. But we found the adsorptive processes were consistent with the non-linearized PFO model. Similarly, Zhang et al. [150] reported GO was fully blended into cellulose matrix and subsequently cross-linked to form the GOCB composites. The removal processes behaved in the Langmuir-type manner and PSO kinetic model. Integrating the merits of the GO and cellulose, the composites turned out with high removal efficiency (MG, >96%) and easy reusability (>5 times), strongly dependent on encapsulation amount of GO, temperature and solution pH.

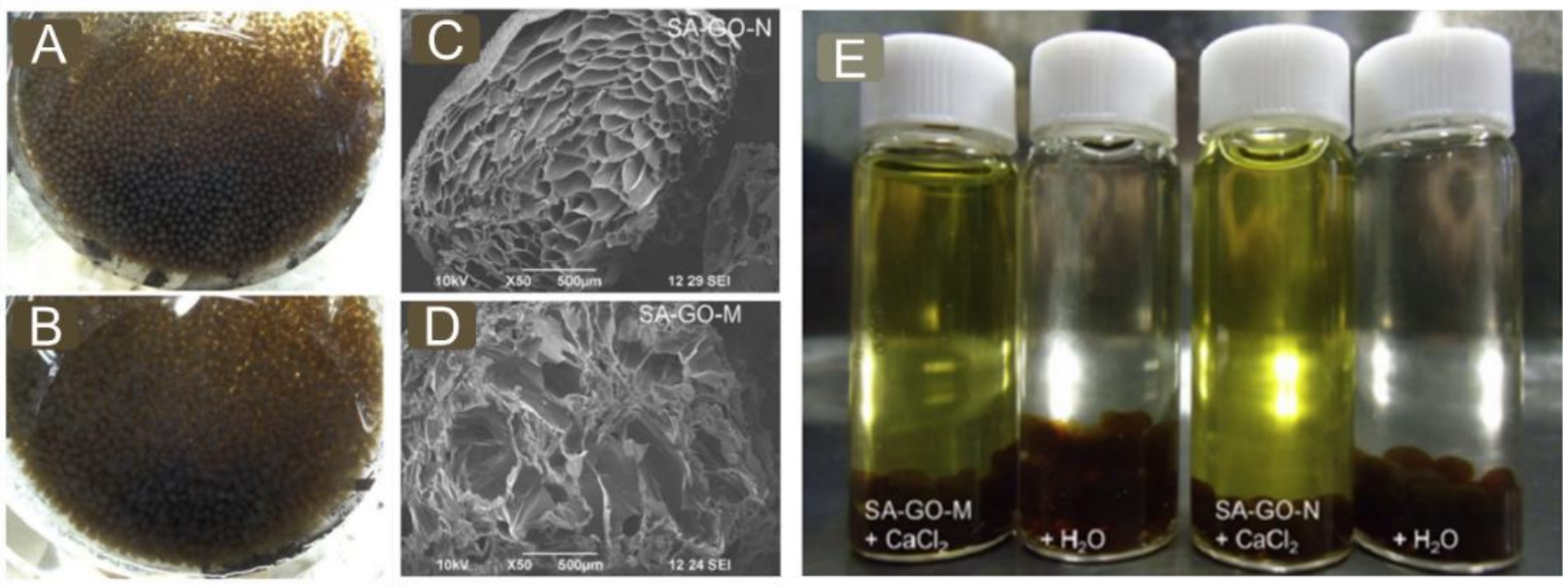

**Figure 10**. (A)$Ca^{2+}$ cross-linked GO-alginate composite beads(SA-GO-N); (B) Acid-gelled GO-alginate composite beads (SA-GO-M);(C) a SEM image of freezing-dry bead SA-GO-N; (D) a SEM image of the bead SA-GO-M; (E) naked-eye comparison of color changes upon dye-adsorbing beads (SA-GO-M and SA-GO-N) were immersed in the $CaCl_2$ ~0.5 mol/L solution and deionized water. Adapted with permission from [121]. Copyright Elsevier, 2014.

Gan et al. presented a konjac glucomannan (KGM) as GO matrix, and prepared GO/KGM based calcium oxide cross-linked hydrogel adsorbent to adsorb MO and MB[139]. Likewise, the presence of GO endowed the hydrogel adsorbent with enhanced adsorbing performance compared to neat KGM hydrogel. And the adsorption followed a PSO kinetics and Freundlich-type multilayered model. Recently, Sabzevari et al. [155] pointed out that current GO framework products were of limited applications in water treatment because of scalability as the result of repulsive hydration forces between GO layers. To this end, they tried to cross-link GO within chitosan to yield a composite (GO-LCTS) via the interaction between the amine groups of chitosan and the carboxyl groups of GO, resulting in enhanced surface area and structural stability. Such changes and the variable morphology of GO-LCTS (402.6 mg $g^{-1}$) were proved to uptake more MB over pristine GO (286.9 mg $g^{-1}$) and in a Langmuir-type adsorption behavior.

In addition to the natural biomaterials, polymer materials are also readily available for this purpose. Actually, the use of polymers to immobilize GO could even allow unique properties for solid phase separation and adsorption as the examples. Huang et al. [133] developed a hypercrosslinked porous polymer monolith adsorbent (polyHIPEs/GO) by modifying the GO with PVP and then hybridizing such nanocomposite into polymeric high internal phase emulsions (HIPEs) further with a high temperature polymerization. Its adsorptions towards cationic MB and RB complied with PSO kinetic model. A further amination of polyHIPEs/GO realized the anionic dye removal, and such composite turned out with a largely improved adsorption capacity ~1967.3 g $g^{-1}$ on the model dye eosin Y. They also noticed that increasing GO in quantity can enhance to some extent both the electrostatic interaction and pi–pi interaction, resulting in a higher uptake of cationic and anionic dyes. From these studies we found even though GO is faded to a minor component of an adsorbent, its naturally occurring multiple interactions with target pollutants still exist and moreover, the aspect of content ratio still influences the final performance of composite adsorbents.

***Morphology-controlled composites: GO as the catalysts template***

In addition to the direct adsorption effect on the various contaminants, the emergence of GO has received extensive attention and shows great potential as additional yet determinative agent/template in the formation of photocatalyst single crystals. During this process, GO can selectively control the growth direction of inorganic nanoparticles owing to the intrinsic hydroxyl, epoxy and carboxylic functional groups, which could act as active anchoring sites for the heterogeneous nucleation of metal ions such as $Au^{3+}$, $Ag^{+}$, $Ti^{3+}$, $Zn^{2+}$, $Ca^{2+}$, etc.[161-163]. Wang et al. prepared new cubic Ag@AgX@Graphene (X = Cl, Br) nanocomposites by a GO sheet-assisted assembly protocol, where GO sheets act as a novel amphiphilic template for hetero-growth of AgX nanoparticles (Figure 11 A). Morphology transformation of AgX nanoparticles from sphere to cube-like shapes was accomplished by involving GO (Figure 11 B). Functional groups containing the hydroxyl, epoxy and carboxylic groups could act as active anchoring sites for heterogeneous nucleation of $Ag^+$ ions [164]. The resultant cubic Ag@AgX@Graphene (X = Cl, Br) nanocomposites exhibit enhanced adsorption capacity, and reinforced electron-hole pair separation in the decomposition of AO dye (Figure 11 C). After that, tetrahedral $Ag_3PO_4$ crystals with morphology transformation from rhombic dodecahedrons has been also confirmed by the hybridization with GO sheets [165]. GO incorporated polymers such as poly (vinyl alcohol) (PVA) has been studied by our group. The role of GO contribute not only the enhancing the mechanical properties but also adjusting the morphology and the porous structure [166,167]. During the hybridization, GO or graphene for this kind of photocatalyst needs to be arranged in a single-layered and/or few-layered manner, otherwise it is not possible to obtain the enhancing efficacy[152,168].

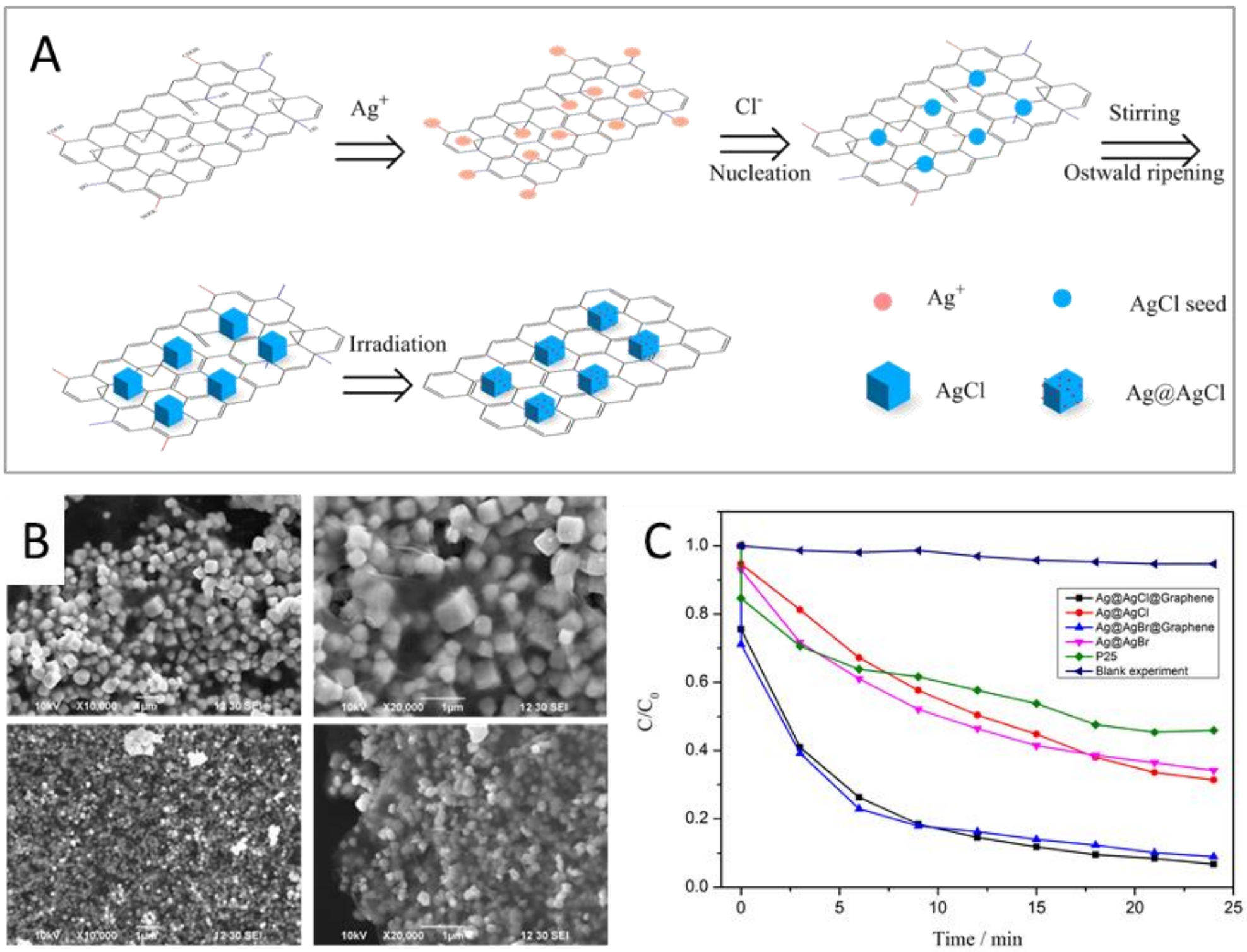

**Figure 11.** Schematic explanation for morphology evolution of the representative Ag@AgCl@Graphene nanocomposites (A). SEM images of the as-prepared cubic Ag@AgCl and quasi-cubic Ag@AgBr nanoparticles encapsulated by gauze-like graphene sheets in Ag@AgCl@Graphene (B, upper) and Ag@AgBr@Graphene (B, lower), respectively. Photocatalytic performance of the thus-prepared Ag@AgX@Graphene plasmonic photocatalysts for the degradation of the AO pollutant under sunlight irradiation (C). Reproduced with permission from [164]. Copyright Royal Society of Chemistry, 2013.

## 4. Prospective outlooks and conclusions

In spite of the enormous progress already achieved in preparation, composited processing and applications of the low-dimensional carbon nanomaterials e.g. carbon nanotubes, graphene oxide and its derivatives within the areas of environmental remediation, challenges and opportunities in practical application remain to be created and grasped. Although many technologies highlighted in this review are still in laboratory level, some of them have undertaken to begin the in-situ tests or even prepared for the commercialization. Among them, adsorption or integrated adsorption-catalysis system would be the most promising way in full scale application based on the cost-effectiveness, safety and operability of carbon nanomaterials. We can anticipate that more research results in laboratory level can be applied into the real solutions through collaborations in terms of research groups and scientific facilities.

Multiple adsorption mechanisms are acting simultaneously on the surface of both CNTs and GO. The mechanism is always complicated and sometimes variable in accordance with the changes of surface functionalization and environmental test conditions. Meanwhile, the catalysis or catalysis support role of CNTs and GO needs to be studied deeply in the future. Constructions of effective adsorption-catalysis integrative systems are appreciated with the aid of template effect of CNTs and GO nanosheets. It is our faith that technologies involved with novel materials and science will eventually solve the environment and energy problems and bring a sustainable world for all in the near future.

**Author Contributions:** Writing-Review & Editing, Y.Q.W., C.P., A.K.V., L.S., W.C.; Editing & Supervision, Y.Q.W.

**Funding:** Y.Q.W. thanks the support from Nihon Trim Co. Ltd. L.S. thanks the support from Beijing University of Technology (105000546317502, 105000514116002) and Beijing Municipal Education Commission (KM201910005007). W.C. thanks the support from National Science Foundation of China (NSFC) # 21872098.

**Acknowledgments:** Y.Q.W. thanks Prof. Bunshi Fugetsu for his advice.

**Conflicts of Interest:** The authors declare no conflict of interest. The founding sponsors had no role in the design of the study; in the collection, analyses, or interpretation of data; in the writing of the manuscript, and in the decision to publish the results.